\documentclass[aps,amsfonts,reprint,
superscriptaddress,
onecolumn,
notitlepage,
nofootinbib,11pt
]{revtex4-1}

\usepackage{amsfonts}
\usepackage{amsmath}
\usepackage{graphicx}
\usepackage{color}

\newcommand{\comment}[1]{}

\begin{document}

\title[Quantum Brownian motion: A converse Zeno scenario]{Quantum Brownian motion under generalized position measurements:
A converse Zeno scenario}

\author{Luca Magazz\`u}
\address{Institute of Physics, University of Augsburg, Universit\"atsstra{\ss}e 1, D-86135 Augsburg, Germany}
\author{Peter Talkner}
\address{Institute of Physics, University of Augsburg, Universit\"atsstra{\ss}e 1, D-86135 Augsburg, Germany}
\author{Peter H\"anggi}
\address{Institute of Physics, University of Augsburg, Universit\"atsstra{\ss}e 1, D-86135 Augsburg, Germany}
\address{Nanosystems Initiative Munich, Schellingstra{\ss}e 4, D-80799 M\"unchen, Germany}
\date{\today}

\begin{abstract}
We study the quantum Brownian motion of a  harmonic oscillator undergoing a sequence of generalized position measurements. Our exact analytical results capture the interplay of the measurement backaction and dissipation. Here we demonstrate that no freeze-in Zeno effect occurs upon increasing the monitoring frequency. A similar behavior is also found in the presence of generalized momentum measurements.
\end{abstract}

\maketitle

\section{Introduction}
\label{intro}
The quantum Zeno effect (QZE), in its original setting, refers to the hindered decay of a frequently monitored  state of a quantum system~\cite{Misra1977, Home1997}. A related phenomenon, called quantum Zeno dynamics~\cite{Facchi2000,Facchi2002,Facchi2008}, generalizes the QZE to the confinement of the system dynamics within a rapidly monitored subspace of the available Hilbert space.
The opposite phenomenon, the accelerated decay due to repeated measurements at slower rates, which takes the name of anti-Zeno effect, turns out  to be much more ubiquitous under realistic conditions~\cite{Kofman2000}.
The transition from the anti-Zeno to the Zeno regime upon increasing the monitoring rate was investigated in Refs.~\cite{Facchi2001, Maniscalco2007,Koshino2005,Chaudhry2016}.
The QZE was first experimentally observed in trapped ions~\cite{Itano1990} and later, along with the anti-Zeno effect, in  a cold atomic gas~\cite{Fischer2001}. Further realizations are found in a variety of experimental setups~\cite{Nagels1997,Kwiat1999,Balzer2000, Streed2006,Molhave2000,Dreisow2008, Bernu2008,Kakuyanagi2015,Bretheau2015, Slichter2016,Harrington2017}.\\
\indent The archetypal model considered in standard treatments of the QZE~\cite{Home1997} consists of an initially excited system (e.g. an atom) decaying in a continuum of states. The QZE is easily understood by observing that the process displays a short-time survival probability of the form $S(\tau)\simeq 1-k\tau^m$ with $m>1$. Assuming instantaneous measurements, the conditional probability of observing the system still in the excited state after $N$ measurements, taken at small time intervals of duration $\tau=t/N$ in a finite time span $t$, is given by $S_c(t)\simeq[1-k\tau^m]^N$.
For $t$ fixed, the conditional survival probability approaches the value $S_c(t)=1$ in the limit $N\rightarrow \infty$, or equivalently $\tau\rightarrow 0$. Hence, the system remains in the initially excited state as long as the monitoring takes place. The particular limiting case $m=1$, which formally would lead to an exponential decay with a $\tau$-independent decay rate, is excluded. Recently, an analysis of the case in which the monitoring time $t$ scales as a power of $N$, i.e., $t\propto N^{\alpha}$, was carried out finding that the Zeno effect occurs for $0\leq \alpha<1/2$~\cite{Facchi2017}.\\
\indent
Apart from a few exceptions \cite{PrinzZwick2015,Milburn1993,Ruseckas2001}, the measurements, which are essential for the QZE, have been treated as instantaneous interruptions of the otherwise unitary dynamics. That means that the interaction between the measured system and the measurement apparatus  must take place within a time span that is very short compared to all relevant time scales of the unitary  evolution. Barchielli et al. \cite{Barchielli1982,Barchielli1983} found a suppression of the Zeno-typical dynamical freezing for sequences of generalized measurements which can be characterized by a strength that decreases with increasing measurement frequency allowing for continuous measurements with ongoing dynamics. A similar approach, which additionally allows for a finite duration of the measurement, was developed by Ruseckas and Kaulakis~\cite{Ruseckas2001}. While for both the approaches in Refs.~\cite{Ruseckas2001,Barchielli1982} the measurements at subsequent times must be performed with identically prepared measurement apparatuses, Gagen et al. \cite{Milburn1993} proposed a model with a single apparatus permanently coupled to the system on which the measurement is performed. Despite the difference of the physical picture, whether there are as many measurement apparatuses as individual measurements, or just a single apparatus whose pointer moves with the measured observable,  the time evolution of the system density matrix is governed by a Markovian master equation of Lindblad type provided that  the measurement strength is properly adjusted to the measurement frequency~\cite{Barchielli1983,Milburn1993}. \\        
\indent Fearn and Lamb~\cite {Lamb1992} analyzed the effect of repeated position measurements of fixed strength on the dynamics of a particle moving in a double-well potential finding a delocalization rather than a freezing of the dynamics in the well in which the particle was initially prepared. This result was challenged in Ref.~\cite{Home1993} claiming that the freezing of the dynamics in either well would result if only sufficiently many measurements within a fixed duration of time were made. Altm\"uller and Schenzle \cite{Schenzle1994} argued differently saying that a proper and more microscopic description of the measurement process would lead to the Zeno effect. An important aspect distinguishing their treatment from those in Refs.~\cite{Lamb1992} and~\cite{Misra1977} is that not a series of measurements with independent measurement apparatuses  but rather a continuous interaction with the electromagnetic field  is considered.
Actually, the reduction of the full two-well system to a two-level system is the feature which enforces the appearance of the Zeno effect in Ref.~ \cite{Schenzle1994}. This is demonstrated by Gagen et al. \cite{Milburn1993} for a model of continuous measurements of the particle position. Indeed, when all energy levels are taken into account\footnote{The validity of the two-level approximation for a double-well system is discussed Ref.~\cite{Magazzu2015}} the localization in the initial well persists for a time span that becomes  smaller with decreasing energy gap between the first two levels and with  increasing measurement  strength. At large enough times a delocalization is always observed.\\
\indent Generally, in a system with a finite-dimensional Hilbert space a sequence of  measurements for detecting the presence of the particle in the initial state yields, at large times, a maximally mixed state of uniform  population~\cite{PrinzZwick2015, Yi2011}. This occurs as long as the time between subsequent measurements is finite. For a vanishing inter-measurement time, however, the Zeno effect manifests itself in its originally proposed form of hindered decay~\cite{Misra1977,Home1997}: The rate at which the asymptotic uniform state is approached becomes zero. This behavior has also been found in the presence of an environment~\cite{Chaudhry2017}, where the details of the dynamics depend  on the specific spectral density of the environment as well as on the strength of the coupling between the system and its environment. \\
\indent With the present work we consider a quantum harmonic oscillator interacting with a \emph{heat bath} of mutually independent quantum harmonic oscillators. The resulting dissipative system provides a model of quantum Brownian motion~\cite{Grabert1984,Grabert1988,LNP484,Hanggi2005}.
The onset of the QZE for quantum Brownian motion has been predicted in Ref.~\cite{Maniscalco2007} within a perturbative treatment starting from 
an exact time-convolution-less master equation~\cite{Schramm1985, Haake1985, HPZ1992, Zerbe1995}. In Ref.~\cite{Maniscalco2007}, the central harmonic oscillator is considered to be initially in a Fock state whose decay is monitored, so that the measured observable, namely the excitation number $\hat{n}$, commutes with the oscillator's Hamiltonian. In contrast, here we investigate the case  of an oscillator which is instantaneously monitored by a so-called Gaussian meter, which measures its position within a certain width, in the same spirit of Refs.~\cite{Milburn1993,Lamb1992}. Notably, such repeated position-like measurement on a harmonic oscillator are of experimental relevance, as  for example in nanomechanical resonators~\cite{LaHaye2004,Giovannetti2004}.\\
\indent In this work we consider the following protocol. We start from the canonical thermal state of the full interacting oscillator-bath system. A first \emph{selective} Gaussian measurement at $t=t_0$ then prepares the oscillator in a state centered at some position $x_0$ with a width $\sigma$. The system subsequently evolves, undergoing a sequence of $n$ \emph{nonselective} measurements; i.e., the measurements leave the system state in a probabilistic mixture of its possible outcomes \cite{Wiseman2010,Watanabe2014}. This intermediate monitoring is performed by Gaussian instruments of width $\sigma$ acting at equally spaced times $t_{j}=t_{j-1}+\tau$, where $j=1,\dots,n$. A final selective  measurement at time $t_F<t_n + \tau$, 
performed with a Gaussian meter, again of width $\sigma$, and centered at $x_F$, provides the two-point probability distribution $W^{(n)}(x_0,t_0;x_F,t_F)$. A scheme of this protocol is provided in Fig.~\ref{scheme} below. We exactly solve the monitored dynamics capturing the interplay between measurements and unitary dynamics of the oscillator interacting with a heat-bath.
In agreement with the finding of Fearn and Lamb~\cite{Lamb1992}, no Zeno effect occurs for the oscillator.
Instead, its position distribution initially spreads with increasing  number of measurements. In the absence of friction, i.e. for an isolated oscillator, the spreading continues and the  position undergoes a diffusion process. Since this diffusion takes place in the confining oscillator potential also the energy grows steadily whereby the position measurements fuel this process: Each position measurement suddenly squeezes the position to a narrow range; the concomitant spreading in momentum space subsequently leads to a spreading in position space beyond the width of the antecedent measurement.
In clear contrast, in the presence of friction the supply of energy by the measurement can now be balanced by the amount of dissipation. Consequently,  the asymptotic  position distribution is characterized, after a sufficiently large number of measurements, by a  finite width.
In neither case a freezing of the dynamics, which is the essential feature of the Zeno effect, occurs. This is because the energy spectrum of the (isolated)  oscillator is unbounded from above. Put differently,  no lower limit exists in the inter-measurement time below which the unitary dynamics of the total system cannot take place between measurements.\\
\indent In a recent treatment of the same model \cite{Bandyopadhyay2014} a survival probability was found that depicts the Zeno effect. However, the corresponding analysis was  based on the iteration of the transition probability for a single pair of measurements thereby neglecting the quantum coherences which  build up during the sequence of Gaussian nonselective position measurements.\\
\indent In the following we describe the model and detail the measurement protocol. Then, we derive the main result, namely the probability distribution for the final measurement conditioned on the result of the first preparing measurement. Finally, we illustrate and discuss the obtained results.

\section{Quantum Brownian motion and generalized position measurements}
\label{model}
The monitored quantum system consists of a one-dimensional quantum harmonic oscillator of mass $M$ and bare angular frequency $\omega_0$, with position and momentum operators $\hat{q}$ and  $\hat{p}$, respectively. This central oscillator interacts via the position operator with a quantum heat bath composed of $N$ harmonic oscillators of masses $m_k$, frequencies $\omega_k$, and coordinates $\hat{q}_k$ and $\hat{p}_k$. The total Hamiltonian reads~\cite{Hanggi2005,Ingold1998}
\begin{eqnarray}\label{H}
\hat{H}=\frac{\hat{p}^2}{2M}+\frac{1}{2}M\omega_0^2\hat{q}^2+\frac{1}{2}\sum_{k=1}^N\left[\frac{\hat{p}_k^2}{m_k}+m_k\omega_k^2\left(\hat{q}_k-\frac{c_k}{m_k\omega_k^2}\hat{q}\right)^2\right]\;.
\end{eqnarray}
The Hamiltonian~(\ref{H}) provides a model for the quantum Brownian motion of a particle in a harmonic potential.
The Heisenberg equation for the position operator $\hat{q}$ of the central oscillator has the form of the following generalized Langevin equation~\cite{LNP484,Hanggi2005,Ingold1998} 
\begin{equation}\label{}
M\ddot{\hat{q}}(t)+M\int_{i_{\rm i}}^t dt'\gamma(t-t')\dot{\hat{q}}(t')+M\omega_0^2\hat{q}(t)=\hat{\xi}(t)\;.
\end{equation}
Here, $\hat{\xi}(t)$ is the quantum Brownian force operator which reads explicitly
\begin{equation}\label{BFO}
\hat{\xi}(t)=-M\gamma(t-t_{\rm i})\hat{q}(t_{\rm i})+\sum_{k=1}^N c_k\left\{\hat{q}_k(t_{\rm i})\cos[\omega_k(t-t_{\rm i})]+\frac{\hat{p}_k(t_{\rm i})}{m_k\omega_k}\sin[\omega_k(t-t_{\rm i})]\right\}\;,
\end{equation}
with $t_{\rm i}$ denoting the time origin. The damping kernel $\gamma(t)$ is given by $\gamma(t)=2(M\pi)^{-1}\int_0^\infty d\omega [J(\omega)/\omega]\cos(\omega t)$, where $J(\omega)$ is the spectral density function defined by $J(\omega):=\pi\sum_k c_k^2\delta(\omega-\omega_k)/(2m_k\omega_k)$.
In the following, if not stated otherwise, we consider a strictly Ohmic heat bath, a bath with spectral density of environmental coupling whose continuum limit is linear in the frequency, i.e.,
\begin{align}\label{Jw}
J(\omega)=\gamma M\omega\;.
\end{align}
The above strictly Ohmic case yields $\gamma(t)=2\gamma\delta(t)$, where the damping parameter $\gamma$ provides an overall measure of the strength of the coupling with the bath modes.\\
\indent We denote by $\rho(t)$  the \emph{total}, time-evolved density matrix for the central and  bath oscillators.  The central oscillator  undergoes a sequence of repeated measurements of its position by  the action of so-called Gaussian meters. A single measurement of the position applied to the full system with density matrix $\rho$ yields a non-normalized post-measurement state of the form~\cite{Ford2007}
\begin{align}\label{GM}
\rho(t)&\rightarrow  f(\hat{q}-x)\rho(t) f^{\dag}(\hat{q}-x)\nonumber\\
&=\int dq dq' f(q-x)f^*(q'-x)\rho_{qq'}(t)|q\rangle\langle q'|\;,
\end{align}
where $\rho_{qq'}(t)=\langle q|\rho(t)|q'\rangle$, $\hat{q}|q\rangle=q|q\rangle$, and where $x$ indicates the center position of the meter. Here, $f(\hat{q})$ denotes a Gaussian slit operator of width $\sigma$,  reading explicitly
\begin{equation}\label{f}
f(\hat{q})=\frac{1}{(2\pi\sigma^2)^{1/4}}\exp\left(-\frac{\hat{q}^2}{4
\sigma^2}\right)\otimes \mathbf{1}_B\;,
\end{equation}
with  the identity operator $\mathbf{1}_B$ acting in the bath Hilbert space.
This Gaussian measurement setting is elucidated in greater detail in~\ref{measurementmodel}. In the limit $\sigma\rightarrow 0$, the measurement action becomes projective, i.e., $\lim_{\sigma \to 0} f(\hat{q}-x) = |x\rangle \langle x|$.
Note that for a finite slit width $\sigma$ the coherences with respect to the position basis are not totally obliterated by the generalized measurement described by Eq.~(\ref{GM}). \\
\indent Starting out  at  a time origin $t_{\rm i}=0$ with the initial density operator of the total system $\rho(0)$, one obtains for the probability density $W(x_0,t_0)$ to find the result $x_0$ in a first position  measurement at some later time $t_0 > 0$  the expression
\begin{eqnarray}
 W(x_0,t_0)&=&{\rm Tr}\{ f(\hat{q}-x_0)U(t_0)\rho(0) U^{\dag}(t_0)f^{\dag}(\hat{q}-x_0)\}\nonumber\\
&=&{\rm Tr}\{f\left (\hat{q} (t_0)-x_0\right )\rho(0) f^{\dag}\left (\hat{q} (t_0)-x_0 \right )\}\:,
\label{W0}
\end{eqnarray}
where $U(t)=\exp(-i\hat{H}t/\hbar)$ is the time evolution operator of the full system, with $\hat{H}$ given by Eq.~(\ref{H}). In Eq.~(\ref{W0}), $\hat{q} (t_0)=U^{\dag}(t_0)\hat{q}U(t_0)$ denotes the position operator in the Heisenberg picture.\\
\indent Similarly, one obtains for the joint probability density $W(x_0,t_0;\dots ;x_F,t_F)$ of finding the central oscillator at the positions $x_0,\;x_1,\;\ldots,\;x_n,\; x_F$ in $n+2$ measurements  at the subsequent times $t_0,\;t_1,\;\dots,\;t_n,\;t_F$ the result
\begin{align}\label{Wk}
W(x_0,t_0;\dots ;x_F,t_F)&={\rm Tr}\{f\left (\hat{q} (t_F)-x_F\right )\dots f\left (\hat{q} (t_0)-x_0\right )\rho(0) f^{\dag}\left (\hat{q} (t_0)-x_0\right )\dots f^{\dag}\left (\hat{q} (t_F)-x_F\right )\}\nonumber\\
&:=\langle f^{\dag}\left (\hat{q} (t_0)-x_0\right )\dots f^{\dag}\left (\hat{q} (t_F)-x_F\right ) f\left(\hat{q} (t_F)-x_F\right )\dots f\left (\hat{q} (t_0)-x_0\right )\rangle\;.
\end{align}

In the following  we assume that the full system is initially prepared at  time $t=0$ in the canonical equilibrium state at temperature $T$, i.e., $\rho(0)=\rho^{\rm th}$, where
\begin{equation}
\label{rho_0}
\rho^{\rm th}= \frac{\exp(-\hat{H}/k_B T)}{{\rm Tr}\{\exp(-\hat{H}/k_B T)\}}\;,
\end{equation}
with $\hat{H}$ the Hamiltonian in Eq.~(\ref{H}). Then, the brackets in Eq.~(\ref{Wk}) (and in the following) denote the canonical thermal expectation value.
It is convenient to introduce the corresponding $(n+2)$-point characteristic function $\phi(k_0,\dots,k_F)$ defined as the Fourier transform with respect to all positions
\begin{equation}\label{xi}
\phi(k_0,\dots,k_F)=\int dx_0dx_1\dots dx_F\; W(x_0,t_0;\dots ;x_F,t_F)\exp{\Bigg(i\sum_{j=0}^Fk_j x_j\Bigg)}\;.
\end{equation}
For the quantum Brownian motion, the characteristic function can be  conveniently cast into the form
\begin{equation}\label{xi2}
\phi(k_0,\dots,k_F)=\prod_{j=0}^F\int dx_j\; f^*(s_j-x_j)f(-s_j-x_j)e^{ik_jx_j}\Bigg\langle {\exp\Bigg(i\sum_{l=0}^F k_l\hat{q}(t_l)\Bigg)}\Bigg\rangle\;,
\end{equation}
where
\begin{equation}\label{gj}
s_j=\sum_{l=j+1}^F k_l\frac{\langle [\hat{q}(t_j),\hat{q}(t_l)]\rangle}{2i}\qquad(j<F)\;
\end{equation}
and $s_F=0$. Details of this derivation can be found in Ref.~\cite{Ford2007}.

\section{Two-point probability distribution with intermediate nonselective monitoring}
\label{Wn}
\begin{figure}[ht!]
\begin{center}
\includegraphics[width=0.6\textwidth,angle=0]{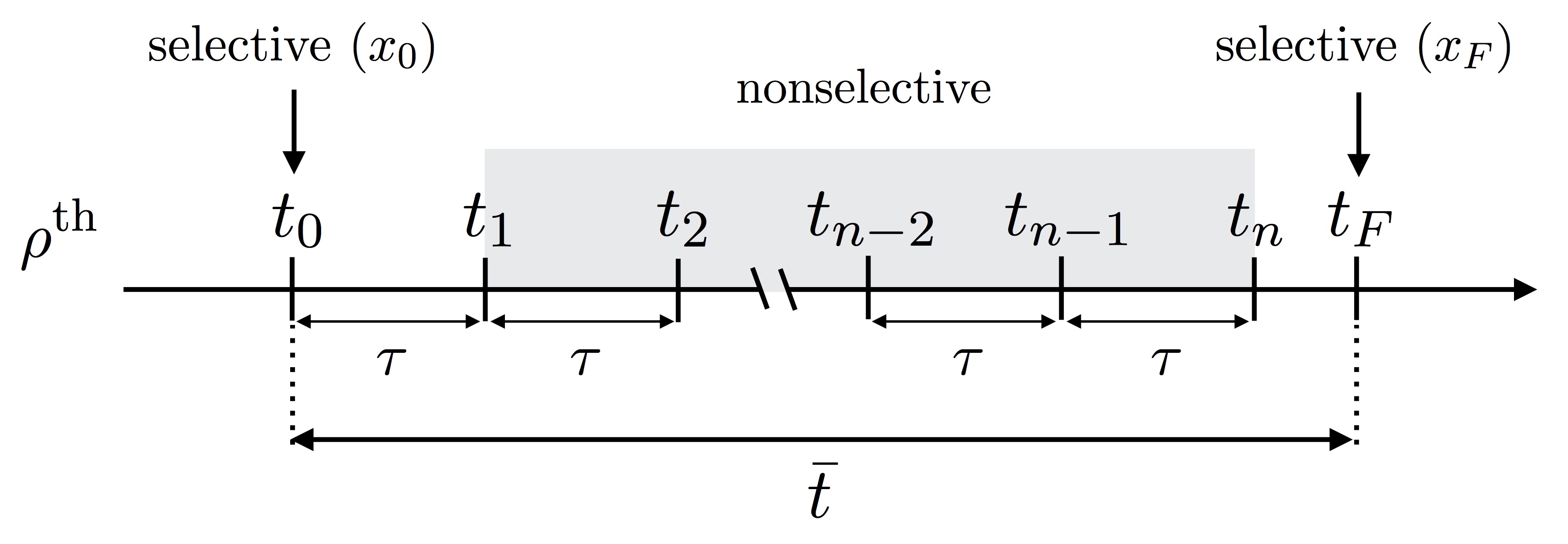}
\caption{\footnotesize{Measurement protocol. The full system described by Hamiltonian~(\ref{H}) is initially in a canonical thermal state. A first selective measurement of the central oscillator position is performed with a Gaussian slit instrument centered at $x_0$ and of width $\sigma$. A sequence of $n$ nonselective measurements by Gaussian slits of the same width takes place with equally spaced measurement times. At time $t_F$, with $t_F-t_n\leq\tau$, a final selective Gaussian measurement with center $x_F$ and width $\sigma$ provides the two-point probability distribution for the position of the oscillator.}}
\label{scheme}
\end{center}
\end{figure}

We next study the situation  in which the $n$ intermediate measurements, with $n\geq 0$, are nonselective, meaning that we integrate the joint $n+2$-point distribution over the $n$ positions $x_j$ of the intermediate Gaussian slits ($j=1,\dots,n$). 
Moreover, we assume that the first $n+1$ measurements occur at equally spaced times $t_j=t_{j-1}+\tau$, where $\tau$ is the time between two successive measurements. A scheme of this protocol is shown in Fig.~\ref{scheme}. 
The resulting two-point probability distribution, with initial and final measurements at times $t_0$ and $t_F$ separated by the total time interval $\bar t:= t_F-t_0 = n\tau + (t_{F}-t_n)$,
is given by
\begin{align}\label{W0F}
W^{(n)}(x_0,t_0;x_F,t_F)&=\int dx_1\dots dx_n\; W(x_0,t_0;x_1,t_1;\dots ;x_n,t_n; x_F,t_F)\nonumber\\
&=\int \frac{dk_0}{2\pi}\dots\frac{dk_F}{2\pi}\;\phi(k_0,\dots,k_F)\int dx_1\dots dx_n\; e^{-i\sum_{j=0}^F k_jx_j}\nonumber\\
&=\int \frac{dk_0}{2\pi}\frac{dk_F}{2\pi}\;\phi(k_0,k_1=0,\dots,k_n=0,k_F)e^{-i(k_0x_0+k_Fx_F)}\nonumber\\
&=\frac{1}{2\pi [\zeta_0^2 \zeta^2( \bar t )-S^2( \bar t )]^{1/2}}\exp\left(\
-\frac{\zeta^2( \bar t )x_0^2-2S( \bar t )x_0x_F+\zeta_0^2 x_F^2}{2[\zeta_0^2 \zeta^2( \bar t )-S^2( \bar t )]}\right)\;,
\end{align}
being a multivariate Gaussian distribution with zero mean, $\overline{ x_0}  = \overline{ x_F} =0$, and with  covariance matrix
\begin{equation}
M \equiv \left ( \begin{array}{cc}
\overline{x^2_0} &\overline{ x_0 x_F}\\
\overline {x_0 x_F}&\overline{ x^2_F}\\
\end{array} \right )
= \left ( \begin{array}{cc}
\zeta^2_0& S(\bar t)\\
S(\bar t)& \zeta^2(\bar t)
\end{array}
\right )
\label{cov}
\end{equation}
determined by the set of quantities
\begin{align}\label{functions}
S(t)&=\frac{1}{2}\langle \hat{q}(t)\hat{q}(0)+\hat{q}(0)\hat{q}(t)\rangle\nonumber\\
\zeta_0^2 &= S(0)+\sigma^2\nonumber\\
\zeta^2( t ) &= \zeta_0^2+\sum_{k=0}^{n}
\frac{A^2( t -k\tau)}{\sigma^2}\nonumber\\
A(t)&=-\frac{i}{2}\langle [\hat{q}(t),\hat{q}(0)]\rangle\;.
\end{align}
The antisymmetrized correlation function in the last line of Eq.~(\ref{functions}) is related to the response function $\chi(t)$ of the central oscillator by $\chi(t)=-2A(t)\Theta(t)/\hbar$. In deriving Eq. (\ref{W0F}) we used the fact that the equilibrium expectation values of the commutator at different times are invariant under time-translations, i.e., $\langle [\hat{q}(t_{F}),\hat{q}(t_j)]\rangle =\langle [\hat{q}(t_{F}-t_{j}),\hat{q}(0)]\rangle$.
Two further comments are in order. First, the two-point probability distribution in Eq.~(\ref{W0F}) carries the label $n$ denoting the number of intermediate measurements. The dependence of the joint probability of the initial and final measurement results on the number of intervening but unregistered measurements is an exquisite feature of quantum mechanics. Put differently, in contrast to a classical stochastic process 
here the quantum measurement's backaction influences the system dynamics, even after tracing over the intermediate measurements.
In the classical limit, the commutators in
Eq.~(\ref{functions}) vanish identically and any reference to the intermediate monitoring is lost.  Second, the complete information about the $n$ unobserved  intermediate measurements entering the joint probability (\ref{W0F}) is contained in the expression $\zeta^2(t)-\zeta^2_0$, as given by the sum in Eq. (\ref{functions}). In the absence of intermediate measurements, this sum reduces to a single term and the wave packet spreading induced by a position measurement (as discussed in Ref.~\cite{Ford2007}) is recovered.\\
\indent The result in Eqs.~(\ref{W0F} - \ref{functions}) holds for any bath spectral density function. Specifically, for the Ohmic bath considered here, the symmetrized and antisymmetrized position correlation functions read~ \cite{Grabert1984,Hanggi2005,Ingold1998}
\begin{align}\label{functions2}
S(t)&= \frac{ \hbar }{ 2 M \omega_r}e^{-\gamma |t|/2} \:\frac{\sinh(\beta\hbar \omega_r)\cos( \omega_r t )+\sin( \beta\hbar \gamma/2)\sin( \omega_r |t| )}{\cosh(\beta\hbar \omega_r)-\cos( \beta\hbar \gamma/2 )}\nonumber\nonumber\\
&-\frac{2\gamma}{M\beta}  \sum_{n=1}^{\infty}\frac{ \nu_n e^{- \nu_n  |t| }}{( \nu_n^2+\omega^2_0)^2- \gamma^2 \nu_n^2}\nonumber\\
A(t) & =  -\frac{\hbar}{2M\omega_r} \sin( \omega_r t )e^{- \gamma |t|/2 }\;,
\end{align}
where $\omega_r =\sqrt{\omega^2_0 - \gamma^2/4}$ denotes the effective frequency of the damped oscillator, $\beta=1/k_BT$ the inverse temperature, and $ \nu_n=2\pi n k_B T/\hbar$  the Matsubara frequencies.  In the limit of vanishing friction  $\gamma\rightarrow 0$ one recovers the results for the free harmonic oscillator prepared in the canonical thermal state. 
%
\section{Results}
\indent  We are interested in the conditional probability density that a measurement taken at time $t_F$ yields the result $x_F$, given that the system was initially prepared by a measurement taken at time $t_0$ with outcome $x_0$, in the presence of $n$ nonselective measurements between $t_0$ and $t_F$, according to the scheme presented in Fig.~\ref{scheme}.
This conditional probability is defined as
\begin{equation}\label{P}
P^{(n)}(x_F,t_F|x_0,t_0)=\frac{W^{(n)}(x_0,t_0;x_F,t_F)}{W(x_0,t_0)}\;.
\end{equation}
The numerator in the rhs. of Eq.~(\ref{P}) is given in Eq.~(\ref{W0F}) while for the denominator we obtain
\begin{align}\label{W0b}
W(x_0,t_0)&=\int dx_F W^{(n)}(x_0,t_0;x_F,t_F)\nonumber\\
&=\int \frac{dk_0}{2\pi}\phi(k_0,k_1=0,\dots , k_n=0,k_F=0)e^{-ik_0x_0}\nonumber\\
&= \frac{1}{(2\pi\zeta_0^2)^{1/2}}\exp\left( -\frac{x_0^2}{2\zeta^2_0}\right) \; ,
\end{align}
 with $\zeta_0^2$ as defined in Eq.~(\ref{functions}) above.\\
\indent Combining  Eqs.~(\ref{W0F}) and (\ref{W0b}) one finds from Eq.~(\ref{P})
\begin{equation}\label{condprob}
P^{(n)}(x_F,t_F|x_0,t_0)=\frac{1}{[2\pi\Sigma_{\tau}^2( \bar t )]^{1/2}} \exp\left\{ -\frac{[x_F-\bar{x}( \bar t )]^2}{2\Sigma_{\tau}^2( \bar t )}\right\}\;,
\end{equation}
which is a Gaussian probability density  function  with mean value
\begin{equation}\label{functions3a}
\bar{x}(t)=x_0\frac{S(t)}{S(0)+\sigma^2}
\end{equation}
and variance
\begin{equation}\label{functions3b}
 \Sigma_{\tau}^2( t )
= \frac{(S(0)+\sigma^2)^2-S^2( t )}{S(0)+\sigma^2} +\frac{1}{\sigma^2}\sum_{k=0}^{n}A^2(t -k\tau)\;.
\end{equation}
Here we made use of the definitions in Eqs.~(\ref{functions}) and~(\ref{functions2}).

\begin{figure}[ht!]
\begin{center}
\includegraphics[width=1\textwidth,angle=0]{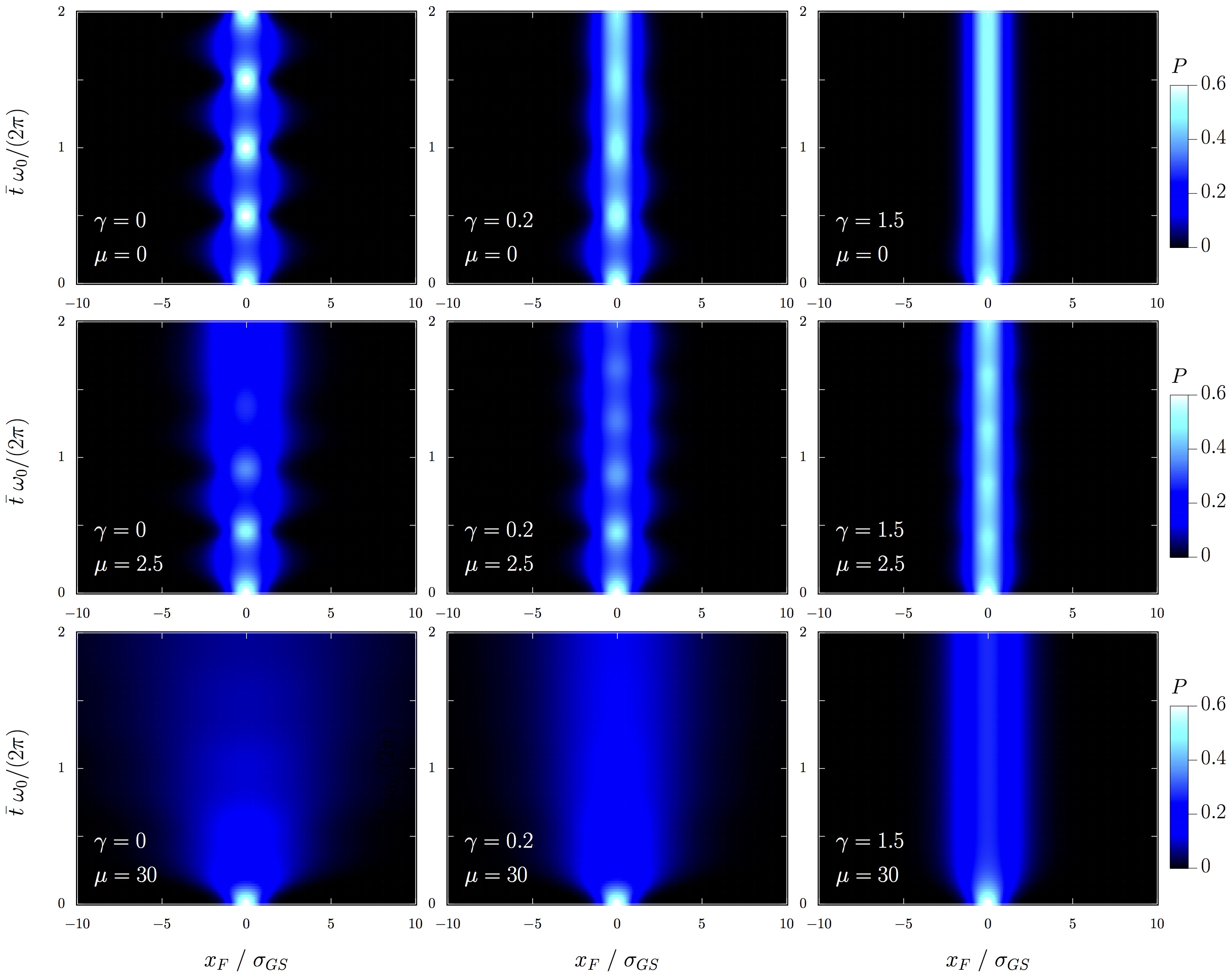}
\caption{\footnotesize{The conditional probability $P\equiv P^{(n)}(x_F,t_F|x_0,t_0)$ [see Eq.~(\ref{condprob})] is displayed as a function of the final position $x_F$ and of the elapsed time $\bar t=t_F-t_0$.
The position $x_F$ is in units of the ground state width of the free oscillator $\sigma_{\rm GS}=(2M\omega_0/\hbar)^{-1/2}$. The friction parameter $\gamma$ increases from left to right and is given  in units of $\omega_0$. The monitoring rate $\mu=\tau^{-1}$ increases from top to bottom and is in units of $\omega_0/2\pi$. 
The measurements are performed according to the scheme in Fig.~\ref{scheme}.
Note the similar behavior at intermediate-friction/intermediate-monitoring-rate (central panel) and strongest-friction/highest-rate (bottom-right panel) reflecting the fact that the measurement-induced spreading is counteracted by the dissipation. At the largest friction value, a low monitoring rate (central-right panel) causes only minor variations with respect to the non-monitored dynamics (upper-right panel), namely a spreading followed by a fast refocusing. The chosen parameters are $x_0=0$, $\sigma=0.5~\sigma_{\rm GS}$, and $T=0.1~\hbar\omega_0/k_B$. The symmetrized correlation function $S(t)$ was numerically determined by truncating the sum in Eq.~(\ref{functions2}) to the first $150$ terms.}}
\label{HO1}
\end{center}
\end{figure}

\begin{figure}[ht!]
\begin{center}
\includegraphics[width=1\textwidth,angle=0]{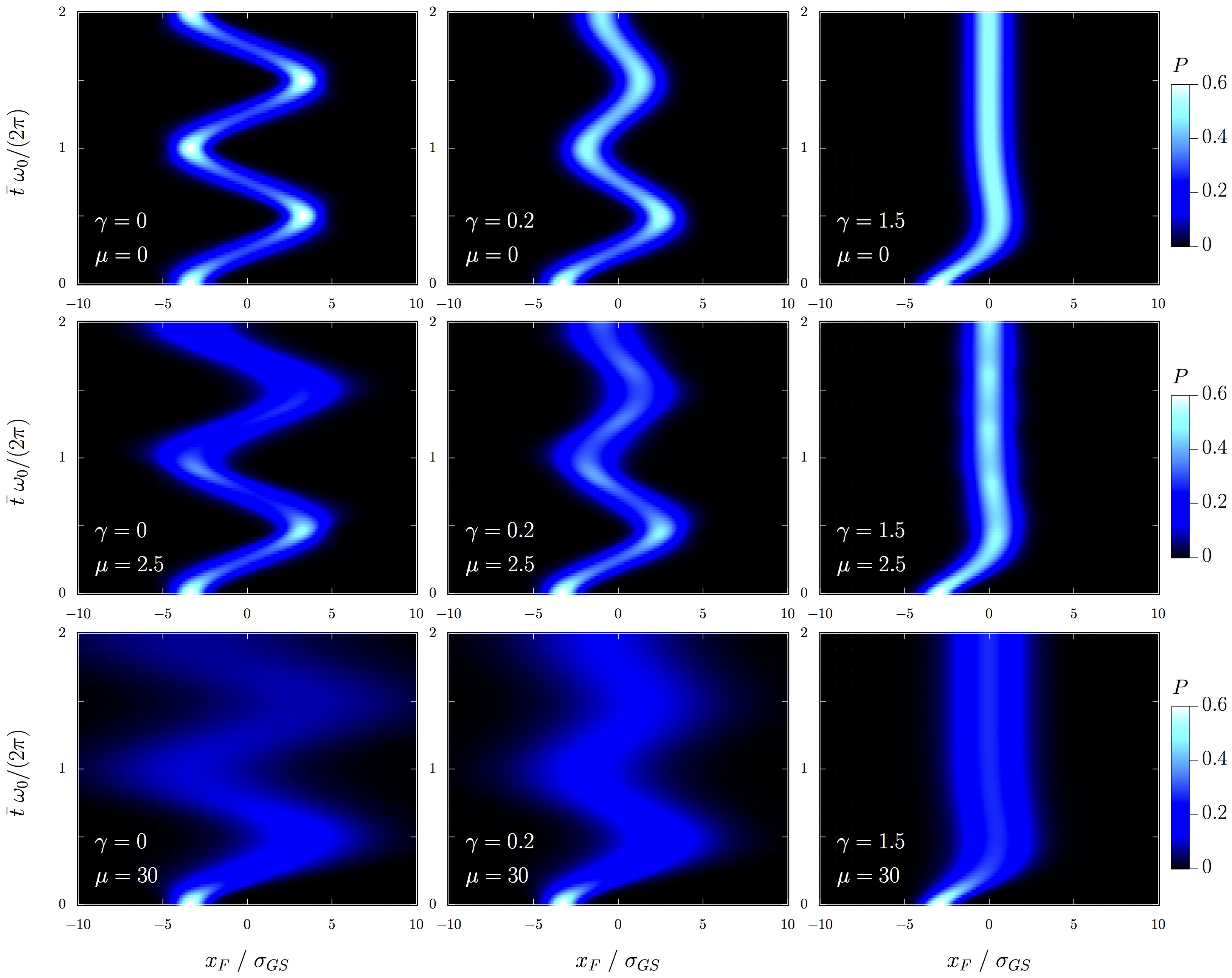}
\caption{\footnotesize{The conditional probability $P\equiv P^{(n)}(x_F,t_F|x_0,t_0)$ [see Eq.~(\ref{condprob})] is displayed as a function of  final position $x_F$ and of time $\bar t=t_F-t_0$. The position $x_F$ is in units of the ground state width of the free oscillator $\sigma_{\rm GS}=(2M\omega_0/\hbar)^{-1/2}$. The friction parameter $\gamma$ increases from left to right and is given  in units of $\omega_0$. The monitoring rate $\mu=\tau^{-1}$ increases from top to bottom and is in units of $\omega_0/2\pi$. Note that the oscillations of the center exclusively depend on the effective frequency $\omega_r$ but not on the presence of intermediate measurements [see Eq.~(\ref{functions3a})].  Here $x_0=-5~\sigma_{\rm GS}$ and the other parameters are as in Fig.~\ref{HO1}. The symmetrized correlation function $S(t)$ was numerically determined by truncating the sum in Eq.~(\ref{functions2}) to the first $150$ terms.}}
\label{HO2}
\end{center}
\end{figure}

\begin{figure}[ht!]
\begin{center}
\includegraphics[width=1\textwidth,angle=0]{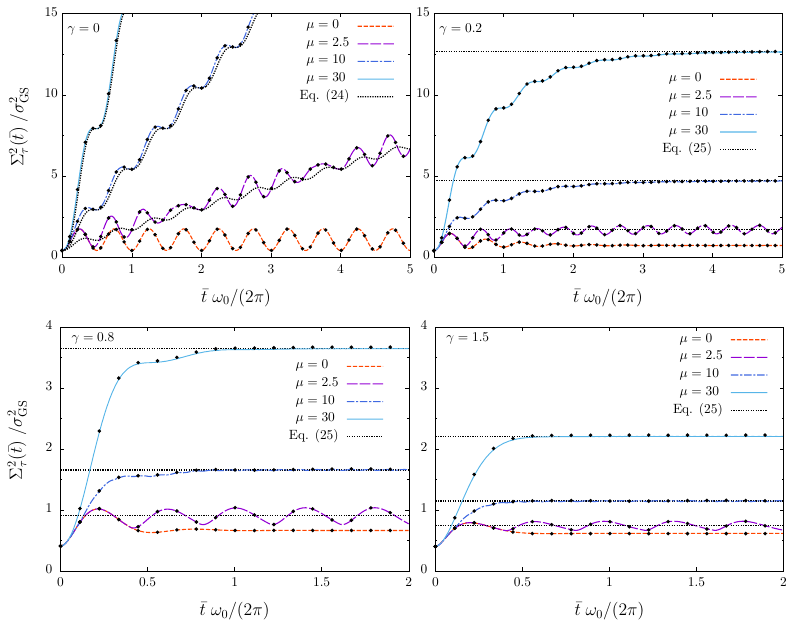}
\caption{\footnotesize{The variance $\Sigma_{\tau}^2(\bar{t})$ [Eq.~(\ref{functions3b})] of the conditional probability density $P^{(n)}(x_F,t_F|x_0,t_0)$, defined in Eq.~(\ref{condprob}), is displayed as a function of the elapsed time $\bar{t}=t_F-t_0$ for  different values of the friction parameter $\gamma$ (in units of $\omega_0$) and of the measurement rate $\mu=\tau^{-1}$ (in units of $\omega_0/2\pi$). The symmetrized correlation function $S(t)$ was numerically determined by truncating the sum in Eq.~(\ref{functions2})  to the first $2000$ terms.  The behavior at large $\mu$ is accounted for by the analytical small-$\tau$ expression~(\ref{sigmasmalltau}). In particular, at $\gamma=0$, the width $\Sigma_{\tau}^2(t)$ increases linearly with time, with a coefficient proportional to $\mu=\tau^{-1}$ [see Eq.~(\ref{limit1})].
On the contrary, for $\gamma\neq 0$, a steady state is reached where the width  gets smaller as $\gamma$ is increased with fixed $\mu$. This is because the larger $\gamma$ the more effectively the energy input from the measurements is dissipated. On the other hand, at fixed $\gamma$, the width increases with $\mu$, as the rate of energy  pumping into the system by the measurement backaction increases [see Eq.~(\ref{limit2})]. \emph{Dotted black lines} -- Predictions from the analytical, small-$\tau$ expressions for vanishing $\gamma$ [Eqs.~(\ref{limit1})] and for finite $\gamma$ at asymptotically large times $\bar{t} $ [Eq.~(\ref{limit2})].  \emph{Diamonds} -- Evaluations for a spectral density function with Drude cutoff at $\omega_D=100~\omega_0$  [see Eqs.~(\ref{SDrude}) and~(\ref{ADrude})]. Other parameters are $\sigma=0.5~\sigma_{\rm GS}$ and $T=0.1~\hbar\omega_0/k_B$}}
\label{Var_q}
\end{center}
\end{figure}

Equations~(\ref{condprob}-\ref{functions3b}) present the main results of this section.
In Figs.~\ref{HO1} and \ref{HO2} the conditional probability density $P^{(n)}(x_F,t_F|x_0,t_0)$ is depicted  as a function of $x_F$ and of time $\bar t$  for three values of the friction strength $\gamma$ and of the monitoring rate $\mu=1/\tau$. The latter parameter is defined as  the number of measurements per unit time and is thus the inverse of the time interval $\tau$.
In both figures $\sigma=0.5~\sigma_{\rm GS}$, where $\sigma_{\rm GS}=(2M\omega_0/\hbar)^{-1/2}$ is the width of the free oscillator's ground state.
In Fig.~\ref{HO1} the condition on the first measurement is specified as $x_0=0$. In the absence of intermediate nonselective measurements (upper row), the dynamics  reflects the motion of the oscillator displaying a periodic spreading and refocusing  in the non-dissipative $(\gamma=0)$ case  (left panel). With increasing friction these pulsations become increasingly damped,  finally leading to a stationary distribution of finite width -- which reflects the motion of a damped harmonic oscillator  (middle and right panels). In the presence of intermediate measurements these pulsations completely disappear. At low measurement rates (middle row) at the very moment of a measurement, the probability density shrinks but expands again a moment after. At higher measurement  rates (bottom row) these indentations become no longer visible and the probability density becomes wider.\\
\indent The same features emerge from Fig.~\ref{HO2}, where for the first measurement we choose $x_0=-5~\sigma_{\rm GS}$. In this case, the average position  follows the motion of a damped harmonic oscillator. The width about the mean value behaves as for the first case with  $x_0=0$.\\
\indent These characteristic features of the dynamical behavior pictured in Figs.~\ref{HO1} and \ref{HO2} can be assessed by inspection of Eqs.~(\ref{functions3a}) and~(\ref{functions3b}), where the mean value and width of the conditional probability distribution~(\ref{condprob}) are given explicitly. First,  using $x_0\neq 0$, then only  in the limit $\sigma\rightarrow 0$ does the average value of $x_F$ at $t_F=t_0$ (i.e. $\bar t=0$) coincide with $x_0$, whereas for a finite width $\sigma > 0$ this average position is found to be $|\bar{x}(0)|<|x_0|$. This is so because the first measurement is performed on the thermal state of the oscillator which has a distribution in position space centered around $x=0$, see Eq.~(\ref{W0b}).
Second, as Fig.~\ref{HO2} indicates, the mean value $\bar{x}( t )$
[Eq.~(\ref{functions3a})] is not affected by the presence of intermediate measurements, contrary to the width of the distribution [cf. Eq.~(\ref{functions3b})]. Indeed, from  Eq.~(\ref{functions}) we deduce that the function $\zeta^2( t )$, which -- as noted above -- exclusively accounts for the effects of intermediate measurements via a series of commutators at different time intervals, only enters  the expression for the width $\Sigma_{\tau}( t )$ of the distribution but not that for the mean value. This observation entails a salient result: Frequent generalized measurements of the oscillator's position  performed with a Gaussian slit apparatus do not hinder the average motion of the system but only affect the spread of the probability distribution. In addition,  the faster the monitoring, the more does the conditional probability  spread, as shown in Figs.~\ref{HO1} and~\ref{HO2}.\\
\indent Finally, it is interesting to study how the variance $\Sigma_{\tau}^2( t )$ of the distribution~(\ref{condprob}) evolves in time and how it is influenced by the monitoring rate $\mu=1/\tau$ as well as by the coupling $\gamma$ to the environment.
In this spirit, the interesting issue  to investigate is whether the traditional Zeno phenomenon eventually emerges for vanishing $\tau$.
For this purpose, assume that the final selective measurement is performed after a time $\tau$ past the last nonselective measurement of the sequence, so that $ \bar t =(n+1)\tau$. Then, considering that $A[ \bar t -(n+1)\tau]=A(0)=0$, the series in Eq.~(\ref{functions3b}) can be approximated by the following time integral in the small-$\tau$ limit
\begin{equation}
\sum_{k=0}^n A^2( t -k\tau)
\simeq
\frac{1}{\tau}\int_0^{t }dt' A^2(t')\;.
\end{equation}
It follows that, in this small-$\tau$ limit, the width of the conditional probability distribution~(\ref{condprob}) emerges as
\begin{equation}
\begin{split}
\label{sigmasmalltau}
\Sigma_{\tau}^2( t )& = \frac{(S(0)+\sigma^2)^2-S^2( t )}{S(0)+\sigma^2}\\
&\quad +\frac{\hbar^2}{8\tau \sigma^2M^2\omega_r^2}\left\{\frac{1-e^{-\gamma  t }}{\gamma}-\frac{\gamma+\left[2\omega_r\sin(2\omega_r t )-\gamma\cos(2\omega_r t )\right]e^{-\gamma  t }}{4\omega_r^2+\gamma^2}\right\}\;.
\end{split}
\end{equation}
This shows that the width of the distribution diverges as $\tau\rightarrow 0$, as it also does for  $\sigma\rightarrow 0$. In the latter limit a projective measurement of position is attained which in turn entails the injection of an infinite amount of energy  upon measuring. \\
\indent From Eq.~(\ref{sigmasmalltau}) two interesting limits of the variance can be taken at fixed, small but finite $\tau$. The first is the frictionless limit at finite time $t$
\begin{equation}\label{limit1}
i)\quad\lim_{\gamma \to 0}\Sigma_{\tau}^2( t ) = \frac{(S(0)+\sigma^2)^2-S^2( t )}{S(0)+\sigma^2}+\frac{\hbar^2}{8\tau \sigma^2M^2\omega_0^3}\left[ \omega_0 t  - \frac{1}{2}\sin(2\omega_0 t ) \right]\;.
\end{equation}
The second is the long-time limit for $\gamma\neq 0$
\begin{equation}\label{limit2}
ii)\quad\lim_{ t  \to \infty} \Sigma_{\tau}^2( t ) = S(0)+\sigma^2+\frac{\hbar^2}{2\tau \sigma^2 M^2\gamma(4\omega_r^2 +\gamma^2)} \;.\qquad\qquad\qquad
\end{equation}
The above limiting cases show that i) for $\gamma=0$, i.e. for an isolated oscillator, the position variance asymptotically spreads proportionally to time under the influence of repeated position measurements and hence the oscillator displays a diffusive behavior in spite of the presence of a confining parabolic potential; 
ii) for $\gamma\neq 0$, i.e., in the presence of an Ohmic  heat-bath, at large $t$, the distribution approaches a finite width which depends on the values of the  parameters $\gamma$, $\tau$, and $\sigma$. Specifically, at fixed $\mu=\tau^{-1}$, the larger the friction constant $\gamma$ the smaller is the variance of the probability density. This behavior is depicted with the six lower panels ($\mu\neq0$) of Figs.~\ref{HO1} and \ref{HO2}. Therefore, in the presence of a heat-bath a stationary regime emerges in which the amount of energy supplied by the position measurements is dissipated to the environment.\\
\indent Further insight into the behavior of the conditional probability density  $P^{(n)}(x_F,t_F|x_0,t_0)$ shown with   Figs.~\ref{HO1} and~\ref{HO2} can be obtained by visualizing the time evolution of its width $\Sigma_{\tau}^2(t)$. In Fig.~\ref{Var_q} this quantity is plotted by using the exact expression~(\ref{functions3b}) for different measurement rates and dissipation strengths $\gamma$. The time evolution and stationary values of the curves at the higher monitoring rates $\mu$ are qualitatively accounted for by the small-$\tau$ limit~(\ref{sigmasmalltau}). In particular, Fig.~\ref{Var_q} depicts the linear increase of $\Sigma_{\tau}^2(t)$ with increasing time at $\gamma=0$~(see Eq.~\ref{limit1}). In contrast, for finite dissipation, we observe a saturating, stationary   behavior, in accordance with  the  analytic expression in Eq.~(\ref{limit2}).\\
\indent We conclude this section with two remarks. First, the results presented are substantially unaffected by the choice of a more realistic Ohmic spectral density function with a cutoff at some finite frequency, as we  checked by using the Drude-regularized position correlation functions provided in~\ref{momentummeasurement}, Eqs.~(\ref{SDrude}) and~(\ref{ADrude}) (see diamonds in Fig.~\ref{Var_q}). Second, the formalism employed does not rely on the particular choice of  the position operator as the monitored observable. Indeed, in~\ref{momentummeasurement} we obtain the  corresponding results for the same measurement sequence but considering instead Gaussian generalized measurements of the oscillator's momentum.

\section{Discussion and conclusions}
\indent With this work we studied the quantum Brownian motion of a dissipative oscillator undergoing a sequence of position-type generalized measurements by so-termed Gaussian slit instruments. The latter  are characterized by a finite width $\sigma$ around a specified position $x$ and yield projective measurements in the limit $\sigma\rightarrow 0$.
The time evolution of the quantum Brownian particle subject to  such repeated, instantaneous measurements was studied through the exact two-point quantum probability distribution with intermediate nonselective measurements. This intermediate monitoring was accounted for by suitably modifying the formalism described in Ref.~\cite{Ford2007}.
We found that an increase of the monitoring rate enhances the position spreading after a first measurement at a given position, the spreading being more dramatic in the limit of vanishing friction. Moreover, the motion of the center of the conditional probability is not affected by the sequence of nonselective measurements.
Hence, none of the characteristic aspects of the Zeno effect are observed at any monitoring rate: Neither is the mean-value affected by the repeated measurements, nor does the quantum state shrink to an eigenstate of the measured observable.
The coupling to an environment leads to the dissipation of the energy injected upon monitoring the system position. In this case, at a large number of intermediate measurements a stationary situation is reached which depends on the dissipation strength. However, the position of the center of the initially prepared state evolves independent of the number of measurements done. The results obtained demonstrate that it is not possible to confine a quantum harmonic oscillator in a certain spatial region by a rapid sequence of instantaneous position measurements, even in the presence of dissipation.\\
\indent There are two clear reasons for this peculiar behavior: The unboundedness of the energy spectrum and the energy increase accompanying each position measurement. 
Because the position as the measured observable does not commute with the system Hamiltonian, the measurement back-action  excites increasingly higher energy states. As a consequence, an initially prepared distribution spreads faster as the frequency of the measurements is increased.
Due to the absence of an upper bound in the energy spectrum, there is no lower limit of time below which the dynamics could freeze. Therefore, even at arbitrarily high monitoring frequencies no traditional Zeno effect does  occur. \\
\indent These results are in contrast to what occurs for projective measurements on systems with bounded Hamiltonians. Under these conditions, 
the conventional Zeno effect follows rigorously \cite{Holevo2001}.    
For a harmonic oscillator under the influence of generalized position measurements both conditions are clearly violated leaving room for a dynamical evolution of the system under permanent observation.


\section*{Acknowledgements}
The authors thank G.-L. Ingold for constructive discussions on this topic. We further acknowledge the support by the  Deutsche Forschungsgemeinschaft (DFG) via the grant HA1517/35-1 (P.H., P.T.) and alike  by the cluster of excellence Nanosystems Initiative Munich (NIM) (P. H.).

\appendix

\section{Modeling the measurement}
\label{measurementmodel}
Following Refs.~\cite{Milburn1993,Talkner2016}, let us model the action of the meter, denoted by $\mathcal{M}$, on the oscillator-bath system described by  Hamiltonian~(\ref{H}). Here, initially, we do not neglect the  evolution of the monitored system during the time $\delta t$ of the system-meter interaction, whose coupling strength $g$ has dimensions of a frequency. The full Hamiltonian is
\begin{equation}
\hat{H}_{\rm tot} = \hat{H} + g \hat{q}\hat{P}_\mathcal{M}\;,
\end{equation}
with $\hat{H}$ defined in Eq.~(\ref{H}).\\
\indent We assume that the interaction starts at $t_0=0$. For $\delta t$ sufficiently small and $g$ not too large, by using the Baker-Campbell-Hausdorff
formula, the time evolution operator can be factorized as
\begin{equation}
U(t) \simeq e^{-\frac{i}{\hbar}\hat{H} \delta t} e^{-iA\hat{q}\hat{P}_\mathcal{M}} e^{-iB\hat{p}\hat{P}_\mathcal{M}}\;,
\end{equation}
where
\begin{equation}
A=\frac{g \delta t}{\hbar}\qquad{\rm and}\qquad B=\frac{g (\delta t)^2}{2M\hbar}\;.
\end{equation}
Note that the operator $e^{ix\hat{P}_\mathcal{M}}$ is a displacement operator for the meter, namely $e^{ix\hat{P}_\mathcal{M}}|Q\rangle=|Q+x\rangle$.\\
Assuming the initial system-meter factorized initial state $\rho_{\rm tot}=\rho(0)\otimes\sigma_\mathcal{M}(0)\equiv\rho_0\otimes\sigma_0$, and that after the time $\delta t$ the meter state is projected into the state $|Q\rangle$~\cite{Milburn1993}, the system state after the measurement reads
\begin{align}
\rho(t)&= {\rm Tr}\left\{|Q\rangle\langle Q| U(\delta t)\rho(0)\otimes\sigma_\mathcal{M}(0)U^\dag(\delta t)  \right\}\nonumber\\
&\simeq\frac{1}{(2\pi\hbar)^2}\int dq d\bar{q}dq' d\bar{q}'dp d\bar{p}\;\sigma_0(Q-Aq-Bp,Q-A\bar{q}-B\bar{p})\nonumber\\
&\qquad\qquad\qquad\times\; \rho_0(q',\bar{q}')e^{\frac{i}{\hbar}p(q-q')}e^{-\frac{i}{\hbar}\bar{p}(\bar{q}-\bar{q}')}|q(\delta t)\rangle\langle \bar{q}(\delta t)|\;,
\end{align}
where $|q(t)\rangle=e^{-i\hat{H}t/\hbar}|q\rangle$.

Now, by neglecting $B$, which is proportional to $\delta t^2$, we get
\begin{equation}
\rho(t)\simeq\int dq d\bar{q}\;\sigma_0(Q-Aq,Q-A\bar{q})\rho_0(q,\bar{q})|q(\delta t)\rangle\langle \bar{q}(\delta t)|\;.
\end{equation}
An additional assumption, which simplifies things further, is that the state of the system under the free evolution induced by $\hat{H}$ alone does not change appreciably during the time interval $\delta t$, i.e., $|q(\delta t)\rangle\simeq|q\rangle$. Whether this assumption is sensible depends on the state of the system previous to the measurement. 
Within the above approximations, the Gaussian measurement is attained by the following choice for the preparation of the density matrix of the meter at the initial time
\begin{equation}\label{A6}
\sigma_0(\xi,\xi')=\frac{e^{-\mu(\xi,\xi')}}{\sqrt{2\pi\langle \hat{Q}^2\rangle}}\;,
\end{equation}
where
\begin{equation}\label{A7}
\mu(\xi,\xi')=\frac{1}{2\hbar^2\langle \hat{Q}^2\rangle}\left\{
\langle \hat{P}^2\rangle \langle \hat{Q}^2\rangle
\left(\xi-\xi'\right)^2-\left[ \langle \hat{Q} \hat{P}\rangle(\xi-\xi')-i\hbar \xi'\right]^2
\right\}\;.
\end{equation}

We then get for the probability density to read from the meter the result $x_0$
\begin{equation}
W(x_0,t_0)\simeq \int dq \;\sigma_0(x_0-Aq,x_0-Aq)\rho_0(q,q)\;.
\end{equation}
This result amounts to taking the trace of the last line of Eq.~(\ref{GM}), provided that $\langle \hat{Q}^2\rangle=\sigma^2$ [cf. Eq.~(\ref{W0})]. Finally, we note that, especially for the non-dissipative case, as the monitoring proceeds and the oscillator is excited to higher energies,
the assumption of instantaneous measurement may break down.

\section{Momentum measurements}
\label{momentummeasurement}
The formalism developed in Sections~\ref{model} and~\ref{Wn} does not rely on the choice of the oscillator's position $\hat{q}$ as the observable being measured. Indeed it also applies to the case in which the measurements are Gaussian momentum measurements with operators
\begin{equation}\label{}
f(\hat{p})=\frac{1}{(2\pi\sigma^2)^{1/4}}\exp\left(-\frac{\hat{p}^2}{4
\sigma^2}\right)\otimes \mathbf{1}_B\;.
\end{equation}
The conditional probability density $P^{(n)}(p_F,t_F|p_0,t_0)$ retains the same structure as the one given in Eq.~(\ref{condprob}) for $P^{(n)}(x_F,t_F|x_0,t_0)$, with the difference is that in Eqs.~(\ref{functions3a}) and~(\ref{functions3b}) one has to use the momentum symmetrized and antisymmetrized correlation functions $S_{pp}(t)$ and $A_{pp}(t)$. These quantities combine to yield the momentum correlation function $C_{pp}(t)=S_{pp}(t)+iA_{pp}(t)$, so that $C_{pp}(0)=\langle p^2\rangle=S_{pp}(0)$. In turn $C_{pp}(t)$ is given by the second time derivative of $C(t)$, namely $C_{pp}(t)=-M^2d^2/dt^2 C(t)$~\cite{Ingold1998}.
As can be seen by inspection of Eq.~(\ref{functions2}), the second time derivative of the symmetrized equilibrium position correlation function $S(t)$ diverges (logarithmically) at $t=0$ in the strictly Ohmic case~\cite{Grabert1984,Talkner1986}.
The physically motivated introduction of a high-frequency cutoff  regularizes this divergent behavior. A simple case is the Drude regularization~\cite{Grabert1984,Karrlein1997,Weiss2012}  for which the spectral density function assumes the algebraically decaying form $J(\omega)= M\gamma\omega(1+\omega^2/\omega_D^2)^{-1}$ [cf. Eq.~(\ref{Jw})], where the Drude cutoff is  $\omega_D\gg \gamma,\omega_0$.
Starting from the expression in Ref.~\cite{Karrlein1997}, after some manipulations, $S(t)$ reads  ($t\geq 0$)
\begin{align}\label{SDrude}
S(t)&=\frac{\hbar }{2M\eta}e^{-\alpha t }\Bigg\{\frac{\delta^2-\alpha^2+\eta^2}{(\alpha-\delta)^2+\eta^2}\frac{\sinh(\beta\hbar\eta)\cos( \eta t )+\sin(\beta\hbar \alpha)\sin( \eta t )}{\cosh(\beta\hbar\eta )-\cos(\beta\hbar \alpha)}\nonumber\\
&-\frac{2\eta\alpha}{(\alpha-\delta)^2+\eta^2}\frac{\sin(\beta\hbar \alpha)\cos( \eta t )-\sinh(\beta\hbar\eta )\sin( \eta t )}{\cosh(\beta\hbar\eta )-\cos( \beta\hbar \alpha)}\Bigg\}
+\frac{2\alpha}{M\beta}\frac{e^{-\delta t}}{\delta(\alpha-\delta)^2+\delta\eta^2}\nonumber\\
&-\frac{2\gamma}{M\beta} \sum_{n=1}^{\infty}\left\{\frac{ \nu_n  e^{- \nu_n  t }}{[\nu_n^2+(\alpha^2+\eta^2)]^2- 4\alpha^2 \nu_n^2}-\frac{ \delta  e^{- \delta  t }}{[\delta^2+(\alpha^2+\eta^2)]^2- 4\alpha^2 \delta^2}\right\}\frac{\omega_D^2}{\delta^2-\nu_n^2}\;.\nonumber\\
\end{align}
The parameters $\alpha$,$\eta$, and $\delta$ are implicitly defied by the relations:
$2\alpha+\delta=\omega_D$,
$\alpha^2+\eta^2=\omega_0^2\omega_D/\delta$, and
$\alpha^2+\eta^2+2\alpha\delta=\omega_0^2+\gamma\omega_D$.
Up to the first order in $\gamma/\omega_D$, the parameters in Eq.~(\ref{SDrude}) read
\begin{align}\label{coefficients}
\alpha &=\frac{\gamma}{2}\frac{\omega_D^2}{\omega_D^2+\omega_0^2}\nonumber\\
\eta &= \sqrt{\omega_0^2-\alpha^2+2\alpha\omega_0^2/\omega_D}\nonumber\\
\delta &= \omega_D-2\alpha\;.
\end{align}
By inspection of Eq.~(\ref{SDrude}) one sees that the strict Ohmic case is recovered in the limit $\omega_D\rightarrow \infty$. Indeed, in this limit, $\alpha\rightarrow \gamma/2$, $\eta\rightarrow \omega_r$, and $\delta\rightarrow\omega_D=\infty$. As a result, Eq.~(\ref{SDrude}) reduces to the corresponding Ohmic expression in Eq.~(\ref{functions2}). On the other hand, for temperatures such that $\nu_n\rightarrow \delta$, the corrsponding $n$-th coefficient in the sum on the last line of Eq.~(\ref{SDrude}) vanishes.\\
The additional contributions to $S(t)$ brought by the Drude regularization [cf. Eq.~(\ref{functions2})] are at most of the order $\omega_0/\omega_D^2$. Nevertheless, the introduction of the cutoff yields for the second derivative of $S(t)$ a non-divergent behavior, which in turn entails the finiteness of $S_{pp}(0)=\langle p^2\rangle$ (see below). \\
\indent Starting from the expression in Ref~\cite{Kumar2014} and using again the relations above involving $\lambda_{1/2}=\alpha\pm i\eta$ and $\lambda_3=\delta$, the antisymmetrized equilibrium position correlation function with Drude regularization can be cast in the form
\begin{equation}\label{ADrude}
A(t)=-\frac{\hbar}{2M\eta}\Bigg\{\frac{2\alpha\eta\left[e^{-\delta t} -\cos(\eta t)e^{-\alpha t}\right]}{(\alpha-\delta)^2+\eta^2}+\frac{(\delta^2-\alpha^2+\eta^2)e^{-\alpha t}\sin(\eta t)}{(\alpha-\delta)^2+\eta^2}\Bigg\}\;.
\end{equation}
Also in this case, with respect to the Ohmic case, the additional terms are of order $\omega_0/\omega_D^2$ at most. Equation~(\ref{ADrude}) reduces to the corresponding Ohmic expression [Eq.~(\ref{functions2})] in the limit $\omega_D\rightarrow \infty$. \\
\indent The symmetrized and antisymmetrized momentum correlation functions are given by $-M^2$ times the second derivative of $S(t)$ and $A(t)$, respectively. We get
\begin{align}\label{SppDrude}
S_{pp}(t)&=\frac{\hbar M}{2\eta}(\eta^2-\alpha^2) e^{-\alpha t }\Bigg\{\frac{\delta^2-\alpha^2+\eta^2}{(\alpha-\delta)^2+\eta^2}\frac{\sinh(\beta\hbar\eta)\cos( \eta t )+\sin(\beta\hbar \alpha)\sin( \eta t )}{\cosh(\beta\hbar\eta )-\cos(\beta\hbar \alpha)}\nonumber\\
&-\frac{2\alpha\eta}{(\alpha-\delta)^2+\eta^2}\frac{\sin(\beta\hbar \alpha)\cos( \eta t )-\sinh(\beta\hbar\eta )\sin( \eta t )}{\cosh(\beta\hbar\eta )-\cos( \beta\hbar \alpha)}\Bigg\}\nonumber\\
&+\hbar M\alpha e^{-\alpha t }\Bigg\{\frac{\delta^2-\alpha^2+\eta^2}{(\alpha-\delta)^2+\eta^2}\frac{\sin(\beta\hbar \alpha)\cos( \eta t )-\sinh(\beta\hbar\eta)\sin( \eta t )}{\cosh(\beta\hbar\eta )-\cos(\beta\hbar \alpha)}\nonumber\\
&+\frac{2\alpha\eta}{(\alpha-\delta)^2+\eta^2}\frac{\sin(\beta\hbar \alpha)\sin( \eta t )+\sinh(\beta\hbar\eta )\cos( \eta t )}{\cosh(\beta\hbar\eta )-\cos( \beta\hbar \alpha)}\Bigg\}
-\frac{2M\alpha}{\beta}\frac{\delta e^{-\delta t}}{(\alpha-\delta)^2+\eta^2}
\nonumber\\
&+\frac{2M\gamma}{\beta} \sum_{n=1}^{\infty}\left\{\frac{ \nu_n^3  e^{- \nu_n  t }}{[\nu_n^2+(\alpha^2+\eta^2)]^2- 4\alpha^2 \nu_n^2}-\frac{ \delta^3  e^{- \delta  t }}{[\delta^2+(\alpha^2+\eta^2)]^2- 4\alpha^2 \delta^2}\right\}\frac{\omega_D^2}{\delta^2-\nu_n^2}
\end{align}
and
\begin{align}\label{AppDrude}
A_{pp}(t)&=\frac{\hbar M}{2\eta}\Bigg\{\frac{2\alpha\eta}{(\alpha-\delta)^2+\eta^2}\left[\delta^2 e^{-\delta t}+(\eta^2-\alpha^2)e^{-\alpha t}\cos(\eta t)-2\alpha\eta e^{-\alpha t}\sin(\eta t)\right]\nonumber\\
&-\frac{\delta^2-\alpha^2+\eta^2}{(\alpha-\delta)^2+\eta^2}e^{-\alpha t}\left[(\eta^2-\alpha^2)\sin(\eta t)+2\alpha\eta \cos(\eta t)\right]\Bigg\}\;.
\end{align}
\begin{figure}[ht!]
\begin{center}
\includegraphics[width=1\textwidth,angle=0]{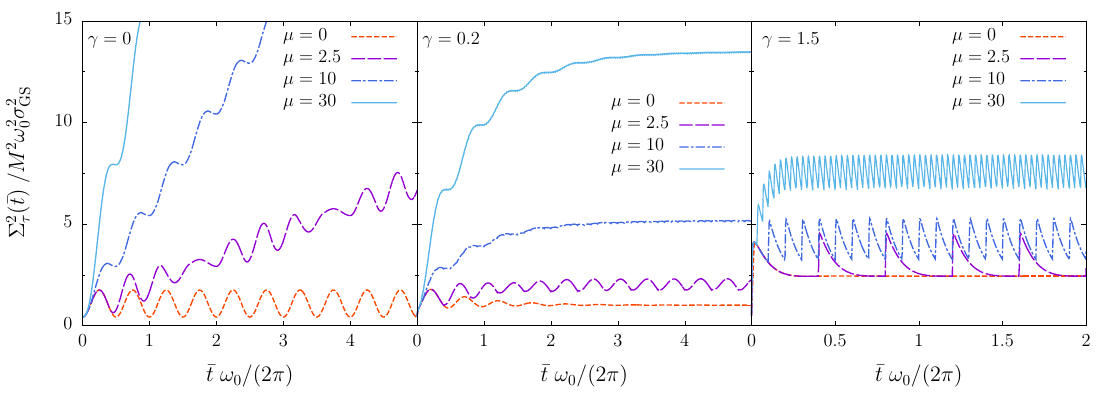}
\caption{\footnotesize{Repeated momentum measurements. The variance $\Sigma_{\tau}^2(\bar{t})$ of the conditional probability density $P^{(n)}(p_F,t_F|p_0,t_0)$ is displayed as a function of the elapsed time $\bar{t}=t_F-t_0$ for four measurement rates $\mu=\tau^{-1}$ (in units of $\omega_0/2\pi$) and for different values of the friction parameter $\gamma$ (in units of $\omega_0$). The latter increases from left to right, taking the same values as in Figs.~\ref{HO1},~\ref{HO2}, and~\ref{Var_q}. The expression for $\Sigma_{\tau}^2(t)$ is given by  Eq.~(\ref{functions3b}) with $S(t),A(t)\rightarrow S_{pp}(t),A_{pp}(t)$.  The symmetrized correlation function $S_{pp}(t)$  was numerically determined by truncating the sum in Eq.~(\ref{SppDrude}) to the first $2000$ terms. 
The qualitative behavior is similar to that predicted for position measurements and shown in Fig~\ref{Var_q}, except that in this case at the highest value of $\gamma$ (right panel) the oscillations induced by the repeated measurements are much more pronounced. Other parameters are $\sigma=0.5~M\omega_0\sigma_{\rm GS}$,  $T=0.1~\hbar\omega_0/k_B$, and $\omega_D=100~\omega_0$.}}
\label{Var_p}
\end{center}
\end{figure}
In Fig~\ref{Var_p} the variance of the probability distribution for repeated momentum measurements is shown. The behavior depicted is similar to that found for position measurements, see Fig~\ref{Var_q}.  Pronounced oscillations around a constant value of the variance at large $\gamma$ constitute the main qualitative difference with the case of position measurements. This different behavior arises because the coupling with the bath oscillators is still via the position operator, whereas the observed quantity is now the momentum.  Beside these minor deviations, the same considerations made about the QZE for repeated position measurements apply for the case considered here.

%

\end{document}